\documentclass{aa}
\usepackage{graphics,natbib,subfig}
\input{psfig.tex}
\begin{document}
%\newcounter{subfigure}
\title{FORS spectroscopy of galaxies in the Hubble Deep Field-South }
\author{Dimitra Rigopoulou\inst{1,2}\thanks{dar@astro.ox.ac.uk}, 
William D. Vacca\inst{3,2},
Stefano Berta\inst{4},\\
Alberto Franceschini\inst{4},
%Michel Dennefeld\inst{3}
and Herv\'e Aussel\inst{5}}
\institute{
\inst{1} Astrophysics, University of Oxford, Keble Rd, Oxford, OX1 3RH, U.K.\\
\inst{2} Max-Planck-Institut f\"{u}r extraterrestrische Physik (MPE),
Postfach 1312, Garching D-85741, Germany \\
\inst{3} USRA, NASA Ames Research Center, MS 144-2 Moffett Field, 
CA 94035-1000 USA\\
\inst{4} Dipartimento di Astronomia, Universita' di Padova,
Vicolo dell'Osservatorio 2, I-35122, Padova, Italy \\
%\inst{3} IAP, France \\
\inst{5} Institute for Astronomy, 2680 Woodlawn Drive,
Honolulu, Hawaii, 96822, USA}
\authorrunning{Rigopoulou et al.}
\titlerunning{FORS spectra of HDFS galaxies}
\date{}
\abstract{
We present low resolution multi-object spectroscopy
of an I-band magnitude limited (I$_{AB} \simeq$ 23--23.5) sample
of galaxies located in an area centered on the Hubble Deep Field-South
(HDFS). The observations were obtained using the
Focal Reducer$/$low dispersion Spectrograph (FORS) on the ESO Very Large
Telescope.
Thirty-two primary spectroscopic targets in the HST-WFPC2
HDFS were supplemented  with galaxies detected in the Infrared Space
Observatory's survey of the HDFS and the ESO Imaging Deep Survey to
comprise a sample of 100 galaxies for spectroscopic
observations. Based on detections of several emission lines, such as
[OII]$\lambda$3727, H$_{\beta}$ and [OIII]$\lambda$5007, or of other 
spectroscopic features, we 
measured accurate redshifts for 50 objects in the central HDFS and
flanking fields. The redshift range of the current sample of galaxies is
0.6--1.2, with a median redshift of 1.13 (at I$\simeq$23.5 not corrected for
completeness).  The sample is dominated by
starburst galaxies with only a small fraction of ellipticals ($\sim$
10\%).  
For the emission line objects, the extinction corrected [OII]$\lambda$3727
line strengths yield estimates of star formation rates in the
range 0.5--30 M$_{\odot} /$yr. We used the present data to derive the
[OII]$\lambda$3727 luminosity function up to redshift of 1.2. 
When combined with 
[OII]$\lambda$3727 luminosity densities for the local and high redshift 
Universe, our results
confirm the steep rise in the star formation rate (SFR) to z$\simeq$1.3.

}
\maketitle

\section{Introduction}
The Hubble Deep Field-South (HDFS, Williams et al.~2000, Casertano et
al.~2000) has been the subject of extensive follow-up ground-based
imaging  observations at optical (ESO-Imaging Survey (EIS), DaCosta et
al.~1998; Big Throughput Camera (BTC) survey, Teplitz et al.~2001),
near-infrared (e.g DaCosta et al.~1998, Vanzella et al.~2001, Rudnick
et al.~2001, Saracco et al.,~2001) and mid-infrared (
Infrared Space Observatory (ISO): Oliver et al.~2002; Mann et
al.~2002; Aussel et al.~2005, in prep.)  wavelengths.  The field has also
been observed by the Spitzer Telescope (Huang et al., 2005, in prep).
The
unprecedented spatial resolution and depth of the Hubble Space
Telescope (HST)  images have allowed a reliable morphological
classification of objects in the field (e.g. the SUNY
collaboration, available at
http:$//$www.ess.sunysb.edu$/$astro$/$hdfs$/$wfpc2$/$wfpc2.html).  In
addition, many groups (e.g. Vanzella et al.~2001, Rudnick et al.~2001)
have used various techniques to estimate photometric redshifts for
large samples of galaxies detected in the HDFS region.  With such a
wealth of imaging observations already in place, a systematic
spectroscopic follow-up of objects detected in the HDFS is imperative
in order to fully exploit the data. To date, only a limited number of
redshifts have been determined for selected objects in the HDFS: Lyman
break galaxy candidates by  Cristiani et al.~2000 and Vanzella et
al.~2002; ISO-detected objects by Rigopoulou et al.~2000 and
Franceschini  et al.~2003; objects in the vicinity of the quasar
J2233-606 by Tresse et al.~1999 and Bergeron et al.~1999.  Recently,
Vanzella et al.~(2002) reported on low resolution spectroscopy of a
sample of 65 galaxies from the HDFS targetting primarily high-redshift
($z > 2$) candidates while Sawicki \& Mallen-Ornellas (2003) targeted 
lower redshift objects.

In this paper, we report on the optical spectroscopy of 100 objects located in 
the HDFS and flanking fields. We have determined accurate redshifts for 50 
objects. The sample selection, observations and data reduction are explained in
Section 2. The redshifts and redshift distribution are given in Section 3
while star formation rates and extinction estimates are discussed in Section 4.
Finally, in Section 5 the [OII]$\lambda$3727 luminosity function and the 
global star 
formation rate densities are presented. Throughout the paper we use 
H$_{0}$= 70 km s$^{-1}$ Mpc$^{-3}$, q$_{0}$ = 0.5 unless otherwise stated.

\section{Sample Selection, Observations, and Data Reduction}

The WFPC2 observations of the HDFS resulted in the detection of a
number of objects down to a limiting magnitude F814W$_{AB} \sim$ 28
mag (Casertano et al.~2000).  From this sample, galaxies were selected
with F814W$_{AB}$ = 23.7 (corresponding to I$_{Vega}$=23 mag), a
limit imposed by the sensitivity of the FORS1 150I grism. This
criterion resulted in a primary sample of 32 objects all within the
central WFPC2 field. The primary sample was supplemented 
with objects from the
Flanking Fields, all of which have identifications in the 
ESO Deep Imaging Survey (EIS) and the BTC catalog. Additionally we 
 included objects selected from the ISO HDFS detections 
(Aussel et al.\ 2005, in prep).
%This resulted in a total sample of 100 objects.
In all, our final sample included 100 objects.

The spectroscopic observations were carried out as part of ESO program
ID 65.O-0418(A) from August 24-28, 2000, using the Focal Reducer
Spectrograph (FORS1) (Appenzeller 2000), in  the
Multi-Object-Spectroscopy (MOS) mode, on  the ANTU--ESO telescope
(formerly UT1), on Paranal, Chile.  The data presented here have been
retrieved from ESO's VLT science archive.

In the FORS1/MOS mode, 19 slits are placed on a mask covering a
field of view (FOV) of 6.8 $\times$6.8 sq. arcmin. To cover the 30
primary objects 5 MOS settings were needed. The supplemental objects
were observed in the slits that were not occupied by the primary
targets in each MOS setting. The slits were 1.2 wide and 22.5
arcsec long. In most cases the target galaxy was placed in the middle
of the slit, so as to enable proper background substraction. In a few
cases a second galaxy happened to fall in the same slit. Spectra of
these serendipitous galaxies were also analysed. The {\it
GRIS-300V}, and{\it GRIS-150I} grisms were used, covering the
4450-8650 \AA\ and 6000-11000 \AA\ ranges, respectively.  The choice
of grism was based on previous photometric redshift
estimates for sample galaxies; for galaxies with z$_{phot} <$ 0.7 the
GRIS-300V grism was used, while the GRIS-150I grism was used for those
galaxies with z$_{phot} >$ 0.7.  The pixel scale was 0.2
arcsec$/$pixel while the resolution was $\lambda / \Delta \lambda$ =
440 and 260 for the {\it GRIS-300V} and {\it GRIS-150I} grisms,
respectively. At each MOS position, three exposures of 1900 s each
were obtained, for a total integration time of 5700 s per position.
The seeing during acquisition of the spectra varied between 0.8--1.5
arcsec.  In Table 1 we present details of the observations.

\begin{table}
%\centering
\caption[]{Log of the FORS1$/$MOS Observations}
\begin{flushleft}
\begin{tabular}{ccccccc}
 & & & & & & \\ \hline
\hline
Field&RA(J2000)&Dec(J2000)&Grism&
Filter&$\lambda / \Delta \lambda$&
$t_{\rm exp}$ (sec)\\ \hline
M 2 &22:32:54.5&-60:33:45.9&300V&GG435&440&5700\\
M 4 &22:33:12.0&-60:32:35.7&300V&GG435&440&5700\\
M 5 &22:32:43.9&-60:33:24.7&150I&GG435&260&5700\\
M 9 &22:32:39.4&-60:33:57.0&150I&OG590&260&5700\\
M 10&22:32:59.6&-60:34:18.1&150I&GG435&260&5700\\
& & & & & & \\
\hline
\end{tabular}
\end{flushleft}
\end{table}

For each field there were 3 separate exposures which were co-added before
spectral extraction.
We reduced the MOS spectra by treating each of the 19 slits separately
and  using standard IRAF\footnote{IRAF is distributed by the National
Optical Astronomy Observatories, which are operated by the Association
of Universities for Research in Astronomy, Inc. under the cooperative
agreement with the National Science Foundation.} routines.  The
individual spectra were bias-subtracted and flat-fielded in the
standard manner.  To remove the background sky emission we fit a
low-order polynomial in the spatial  direction at each column and then
subtracted it for each individual exposure. Most of the cosmic ray
hits in the sky region of the spectral images were excluded from the
fits with a sigma-rejection algorithm. We then extracted a
variance-weighted one-dimensional spectrum.  The final spectrum was
wavelength calibrated using arc lamp spectra and flux-calibrated using
the spectro-photometric standard stars included in the MOS frame.  The
spectra of the standard stars were reduced in the same way as the
galaxy spectra.

In Table 2 we present an inventory of all objects targeted with
FORS1$/$MOS. We list galaxy number, coordinates (in those cases where
two galaxies are included in the slit we report only the position of
the sample galaxy) and optical (UBVRI) and near-infrared (JHK)
photometry. We give AB magnitudes, taken mostly from EIS but
supplemented with data from the BTC survey available from
http:$//$hires.gsfc.nasa.gov$/$research$/$hdfs-btc.  Additionally, we
list HST identifications from the HDFS-WFPC2 catalogue (available from
http:$//$archive.stsci.edu$/$pub$/$hdf-south$/$version2) and the HDFS
WFPC2 flanking fields catalogue (available from http:$//$
archive.stsci.edu$/$pub$/$hdf-south$/$version1). Finally, we also give
ISO identifications (from Aussel et al., 2004 in prep).

HERE, TABLE 2 TO BE INSERTED
\vskip 30mm

\section{Results}

\subsection{Spectroscopic measures}

In Figures 1a-1f and 2a-2b we show
%the sky-subtracted 2-D spectrum together with
the final wavelength- and flux-calibrated spectra for each object in
the M2, M4, M5, M10 (Figs. 1a-1f) and M9 fields (Figs. 2a-2b). The spectra have
been grouped in the two figures according to the wavelength coverage:
spectra in Figure 1 cover the 4450 to 8200 \AA\ range (corresponding
to the GRIS-300V grism) while those in Figure 2 cover  the 6000-9200
\AA\ range (corresponding to the GRIS-150I grism).  Redshifts were
determined based on the wavelengths of the detected emission lines
(usually the [OII]$\lambda$3727 line originating in a starburst)
or, in a few cases,  absorption features such as the CaII H, K
features (at 3968.5   and 3933.7 \AA) prominent in the spectra of elliptical
galaxies.
When multiple emission lines were detected, we averaged the
redshifts from these lines.  In the cases where we detect
absorption features we have not attempted to measure any properties of the
feature.
%For those objects for which we detect continuum but no
%emission$/$absorption lines an arrow indicates the expected position of
%the emission line  if the redshift is known from additional
%observations (such as in the case of ISO detected galaxies).
The shape of the underlying continuum (especially in cases where emission 
lines were present) was also taken into account and helped to
constrain the redshifts more efficiently.
In each spectrum we report the galaxy number (from Table 2) and
the corresponding redshift and mark the position of both detected and expected 
emission$/$absorption lines. 
Note that emission features are not always seen at the expected locations of
some lines (e.g., H$\beta$ or [OIII]$\lambda$5007). 
In several spectra there are residual [O {\sc i}] night-sky
lines at $\lambda
\lambda$ 5577 and 6300-6364   that can sometimes lead to
``spurious'' emission features in the spectra. These lines are highly variable 
and could not always be properly removed during the sky subtraction step.

In total we have obtained spectra for 100 objects. Some of the objects 
were determined to be stars.
We have detected continuum in 78 objects and measured redshifts in 50 of them.
We supplemented our redshift estimates with additional values taken from the literature (Rigopoulou et al.~2000 and Franceschini et
al.~2003).
\begin{subfigures}
\begin{figure*}
\psfig{file=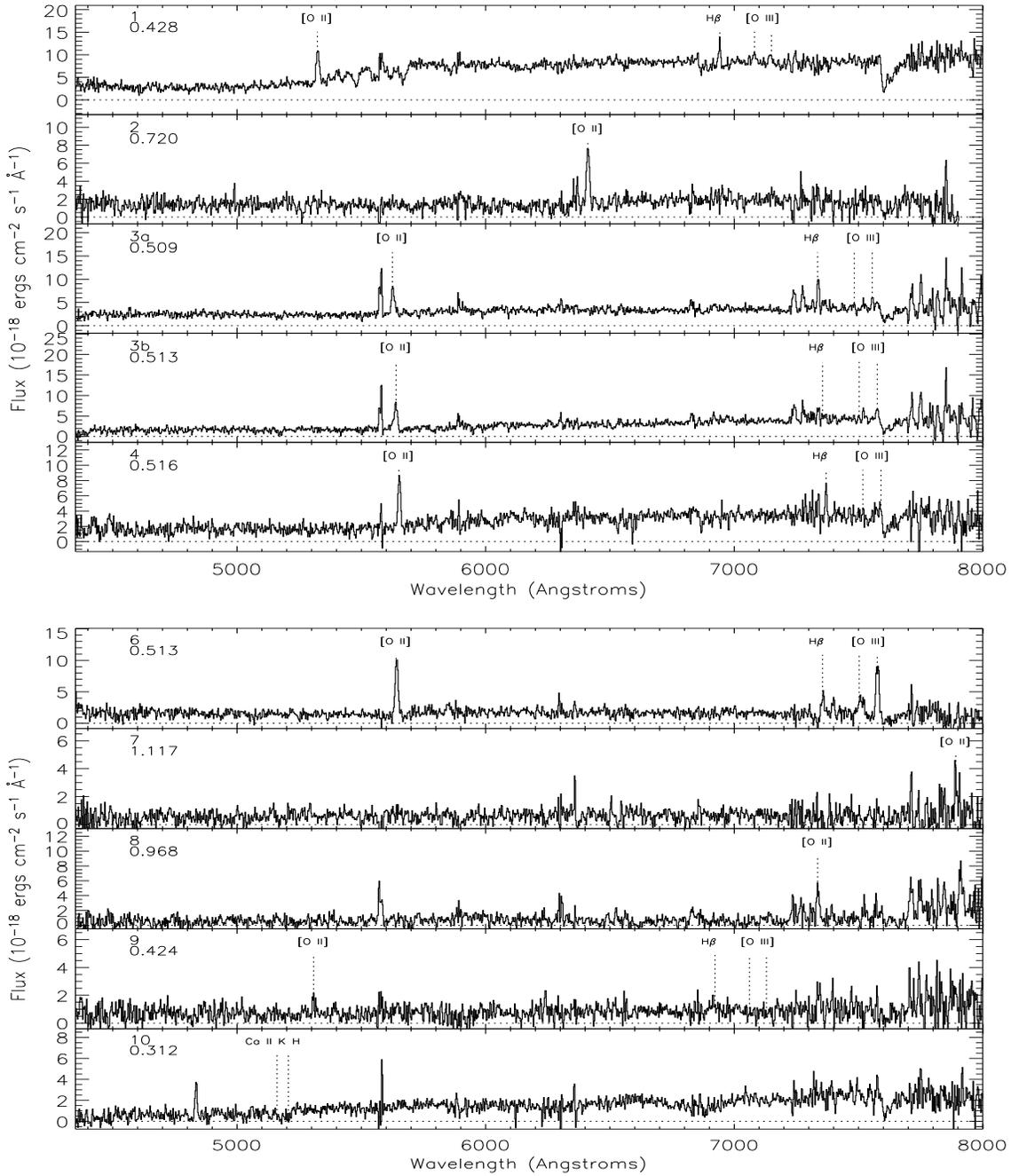,height=23.0 cm,width=18 cm,angle=0}
\caption{Spectra of galaxies observed in the HDFS. The ordinate gives the
flux (in erg cm$^{-2}$ s$^{-1} \AA^{-1}$)). The wavelength coverage extends
between 4200\AA  and  8200\AA. The numbers on the top left corner of
the spectra correspond to objects from Table 2 while the second number
denotes the redshift.  
Note that for objects 1--39 the resolution is $\Delta
\lambda / \lambda \sim$ 440 while for objects 40 to 94 resolution is
$\Delta \lambda / \lambda \sim$ 260 (due to different grism filter
combination). Since the wavelength coverage is common we present both
spectra together.}
\end{figure*}
\begin{figure*}
\psfig{file=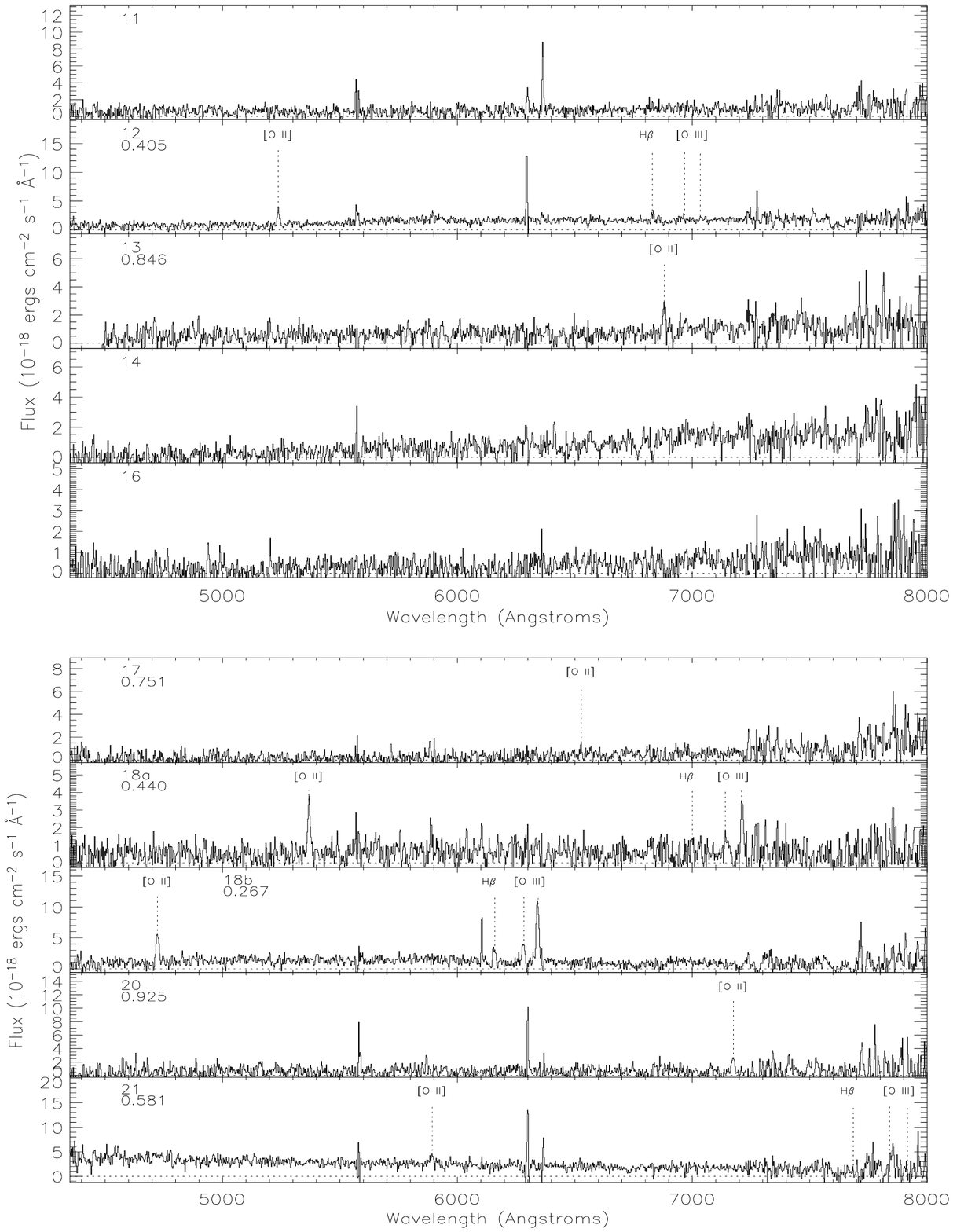,height=25.0 cm,width=18 cm,angle=0}
\caption{cont.}
\end{figure*}
\begin{figure*}
\psfig{file=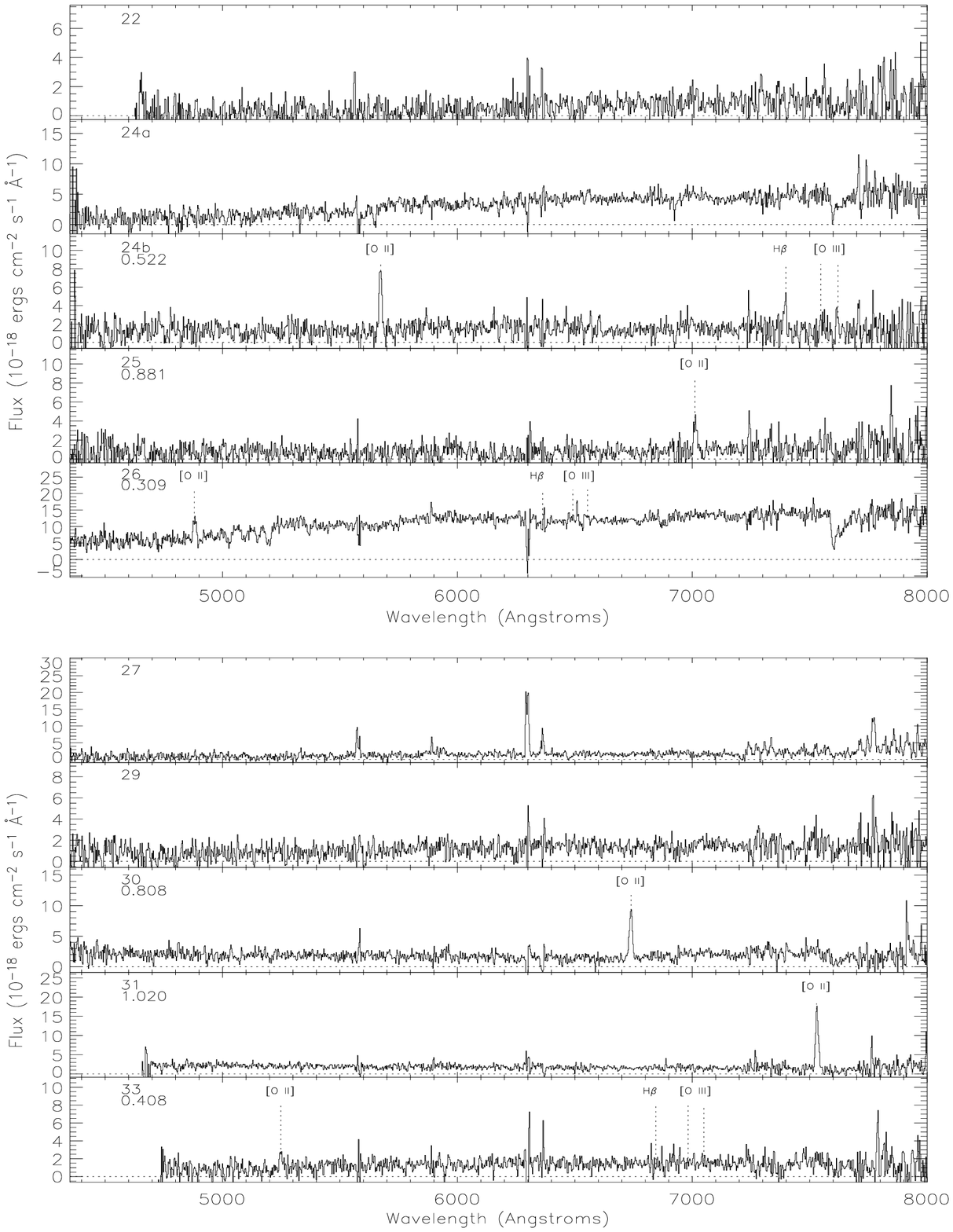,height=25.0 cm,width=18 cm,angle=0}
\caption{cont.}
\end{figure*}
\begin{figure*}
\psfig{file=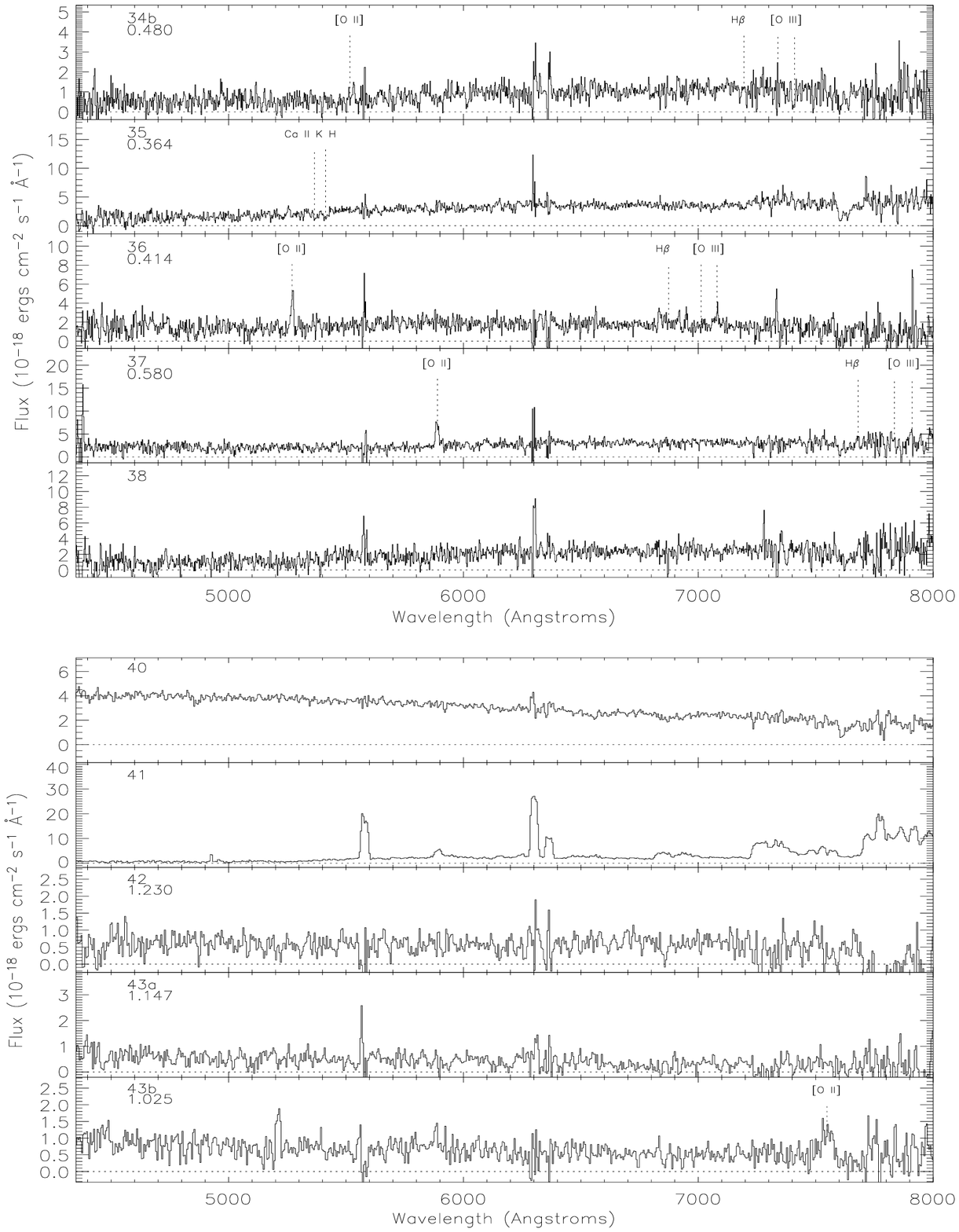,height=25.0 cm,width=18 cm,angle=0}
\caption{cont.}
\end{figure*}
\begin{figure*}
\psfig{file=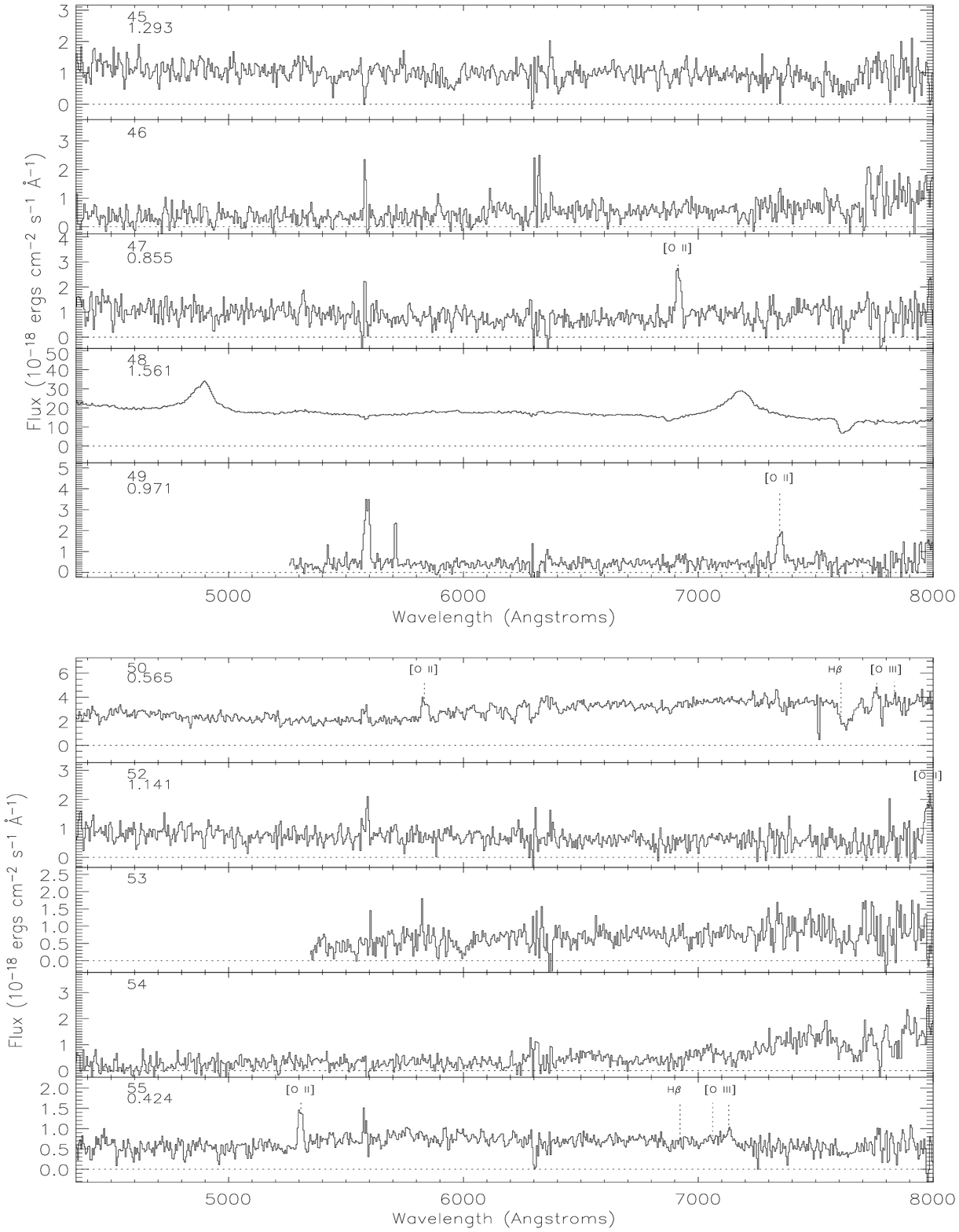,height=25.0 cm,width=18 cm,angle=0}
\caption{cont.}
\end{figure*}
\begin{figure*}
\psfig{file=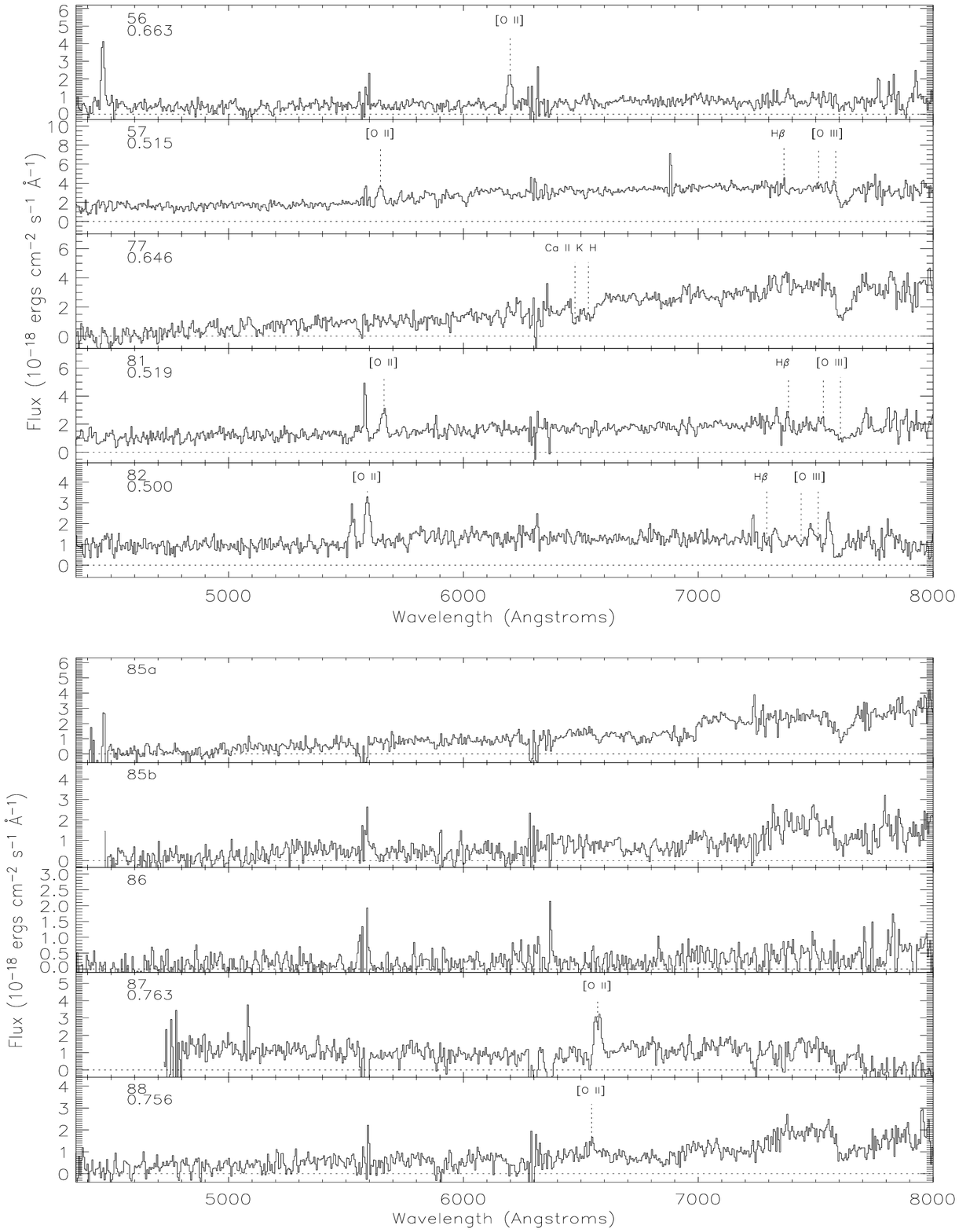,height=25.0 cm,width=18 cm,angle=0}
\caption{ cont.}
\end{figure*}
\begin{figure*}
\psfig{file=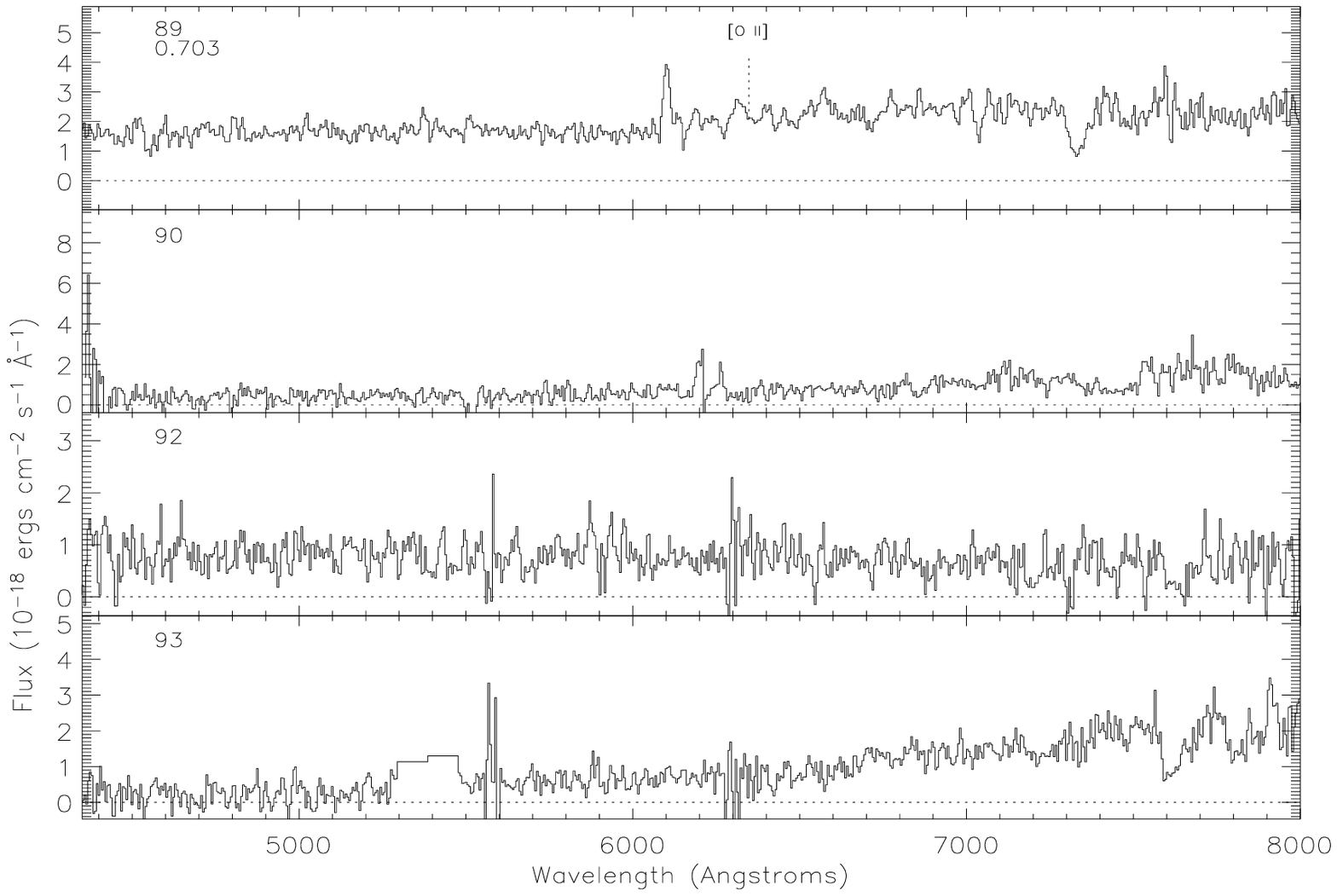,height=25.0 cm,width=18 cm,angle=0}
\caption{cont. }
\end{figure*}
\end{subfigures}

\begin{subfigures}
\begin{figure*}
\psfig{file=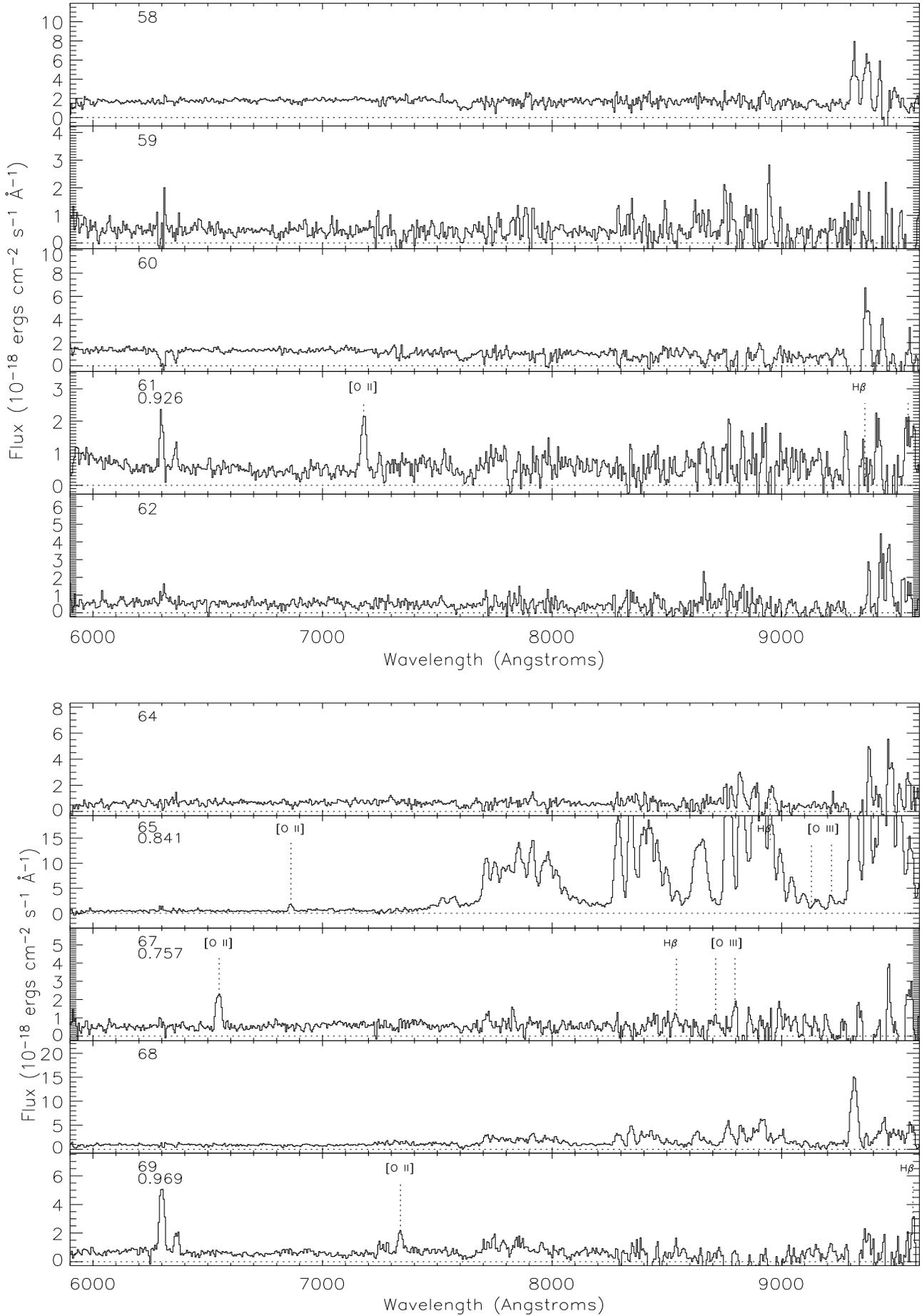,height=25.0 cm,width=18 cm,angle=0}
\caption{Same as Figure 1. The wavelength coverage extends 
between 6000 -- 9200\.A}
\end{figure*}
\begin{figure*}
\psfig{file=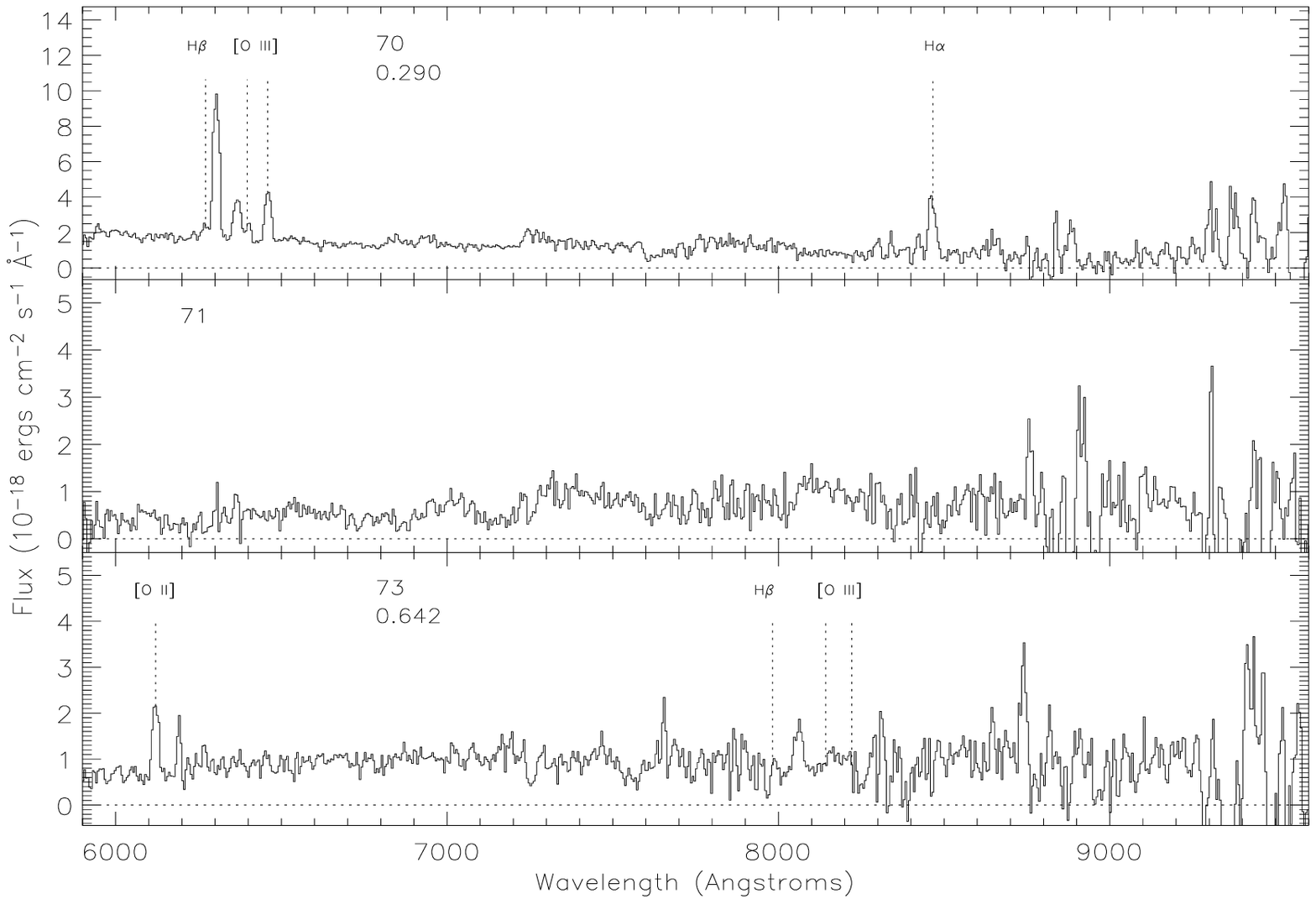,height=10.0 cm,width=18 cm,angle=0}
\caption{cont. }
\end{figure*}
\end{subfigures}

In Table 3 we give the redshifts and the emission line fluxes measured
from the present spectra. In the majority of the spectra we detect
[OII] $\lambda$3727   and in some cases the [OIII] $\lambda$4958  
and $\lambda$5007   lines as well. In half a dozen galaxies we measured
H$_{\beta}$ $\lambda$4861   line fluxes. Our measured redshifts are in 
agreement with the redshifts of Sawicki and Mallen Ornellas (2003), although
the two lists are complementary as these authors target
lower redshift objects.
Although we did not perform a detailed classification of
objects based on their spectral properties, it is evident that the majority
of the sample seem to be starburst galaxies in the redshift range
z=0.6--1.2. We detected only a small fraction (less than 10\%) of
elliptical galaxies \footnote{We labeled as ``ellipticals'' those
galaxies whose spectra show the absorption CaII H, K features.}.
The highest redshift object in our sample (no.\ 48) is a QSO at
z=1.561. Because of its redshift the Balmer lines are redshifted into
the near-infrared regime; in our optical spectra we detected
C {\sc iii}] $\lambda$1909   and MgII $\lambda$2798 \AA. There is very good
agreement between the redshift determined
from the present optical spectra and the near-infrared
spectra of Franceschini et al.~(2003).
Redshifts for objects nos.\ 37 and 89 in our sample
were determined independently from near-infrared data as well, and the 
agreement with our estimates is quite good.
%For those objects where continuum has been detected with a hint of an emission
%line or absorption feature, we place limits on the expected
%redshifts based on the shape of the continuum. 
Based on the measured [OII]$\lambda$3727 line fluxes we
have calculated the individual expected SFR([OII]$\lambda$3727) 
(see Section 4 for details). We note that the spectral profiles were
examined carefully in order to exclude objects displaying broad lines (ie
containing type 1 Seyferts and QSOs) from the further estimates of the 
luminosity function (LF) and
the total SFR density (see section 5.1, 5.2).

HERE, TABLE 3 TO BE INSERTED

\vskip 30mm

\subsection{Median Redshift}

We estimated the median redshift as a function of the I$_{AB}$ magnitude
using the values given in Table 2. A plot of the median redshift vs. 
I$_{AB}$ magnitude is presented in Figure 3. For the plot we have 
used the entire FORS sample, i.e.
objects from the main WFPC2 area, the Flanking Fields
and the ISO selected targets (see Table 2). 
Although we have determined redshifts for the majority
of the objects, the sample is rather incomplete at the fainter magnitudes. 
%We correct for the incompleteness by dividing 
%the number of galaxies with spectroscopic redshifts in each $i$-th magnitude 
%bin by the total number of galaxies in each bin.
Assuming that n$_{i}$(z) and n$_{i}$(tot) are, respectively, the number 
of galaxies with spectroscopic redshift and the total number of galaxies (down
to I$\sim$23.5), then the completeness function in a given magnitude interval
is defined as:

\begin{equation}
\eta_{i} = n_{i} (z) / n_{i} (tot)
\end{equation}

In estimating the completeness function (Eqn. 1) we  
assumed that there are no biases or systematic effects and that the measured
redshifts provide a fair sample of all possible redshifts in each bin.
We then correct for the incompleteness by dividing the number of galaxies 
in each bin by $\eta_{i}$.
In Figure 3 we plotted the galaxy redshift as a function of I$_{AB}$ 
magnitude for the entire FORS sample. We find that a median redshift of 1.16
 is reached by I$_{AB} \sim$ 23.5 (corrected for incompleteness).

\begin{figure*}
\psfig{file=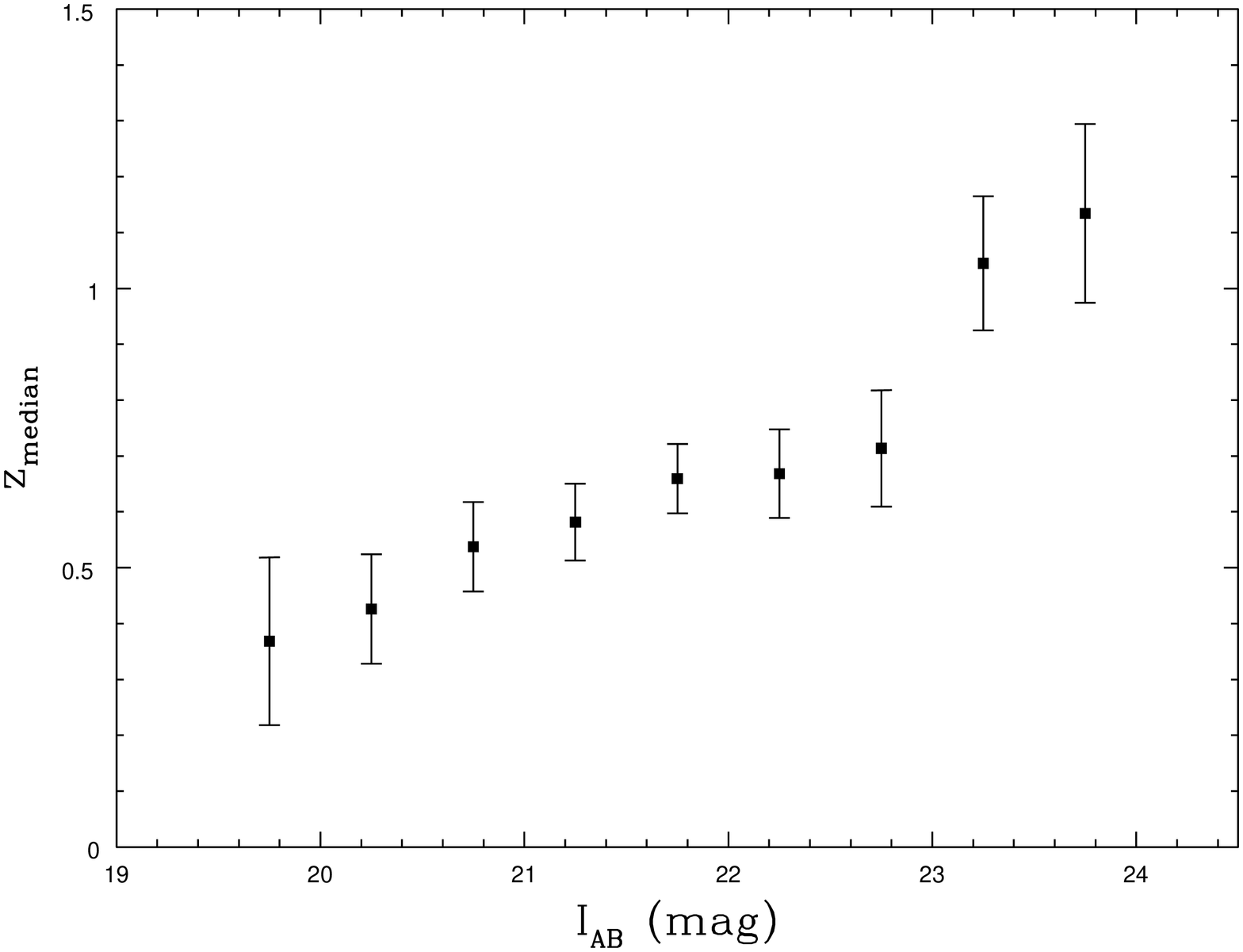,height=16.0 cm,width=16 cm,angle=-90}
\caption{Median redshift as a function of I$_{AB}$.}
\end{figure*}

Finally, Figure 4 shows the redshift distribution for our
spectroscopic sample. The distribution shows an evident peak at
z$\sim$0.58 which is most likely due to the large scale structure present
in the HDF-S region (e.g. Arnouts et al.~2002, Vanzella et al.~2002).

\begin{figure*}
\psfig{file=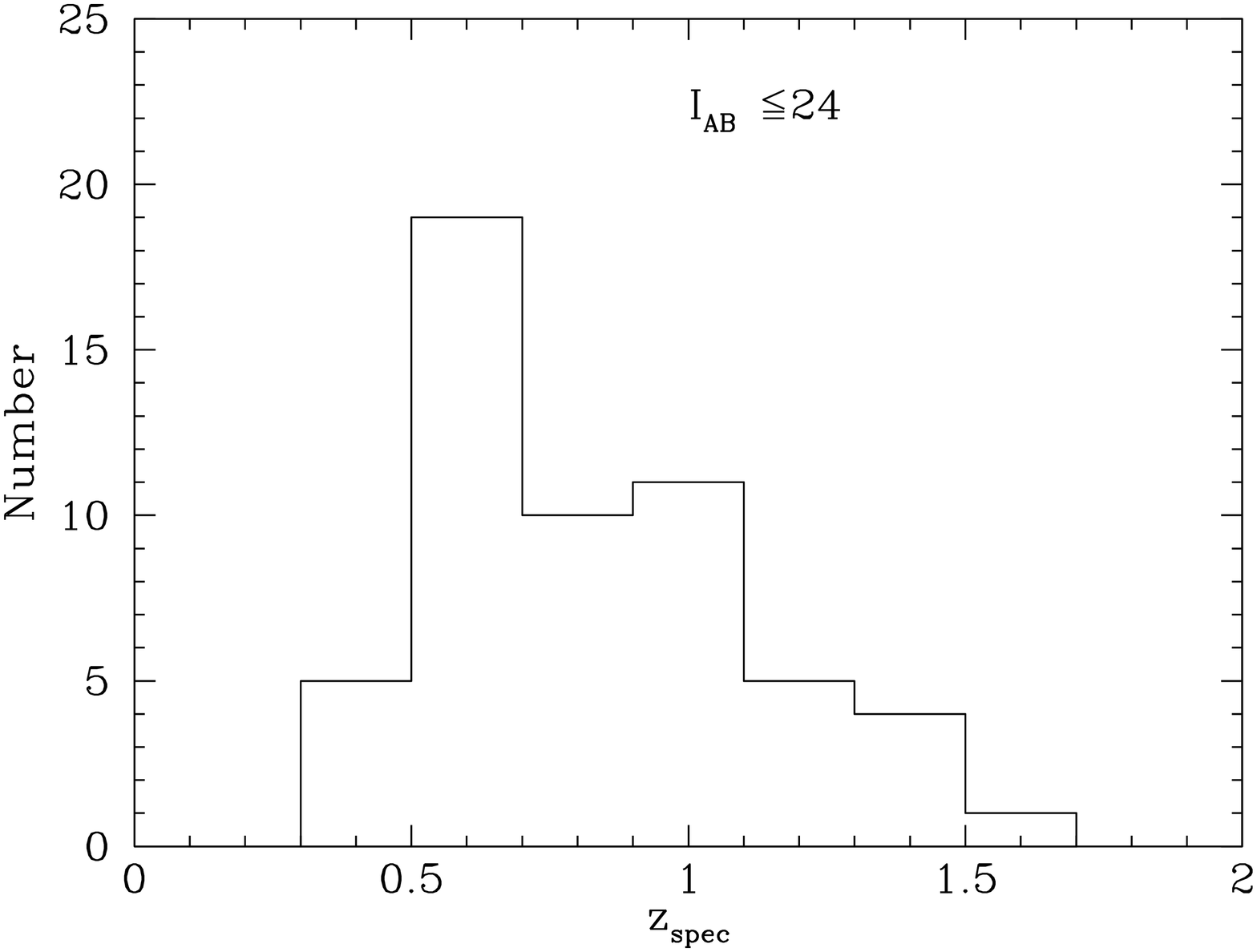,height=16.0 cm,width=18 cm,angle=-90}
\caption{Redshift distribution for all objects with confirmed spectroscopic 
redshifts. }
\end{figure*}

\section{Extinction and SFR estimates}

For ionization-bounded HII regions the Balmer emission line
luminosities scale directly with the ionizing luminosities of the
exciting  stars and thus are proportional to the star formation rate (SFR).
Therefore, it is possible to use the Balmer emission lines to derive
SFR in each galaxy.
Traditionally, H$_{\alpha}$ has been used to estimate SFR in galaxies in the
redshift range up to z$\sim$0.4; beyond this redshift, H$_{\alpha}$ moves 
out of
the optical window and can be accessed at z$>$0.8 in the near-infrared regime.
At redshifts 0.5$<$z$<$1.4, the [OII] $\lambda$ 3727 emission line
can be used to study the star formation rates.

The conversion factor between [OII] luminosity and SFR is computed
using an evolutionary synthesis model. To estimate SFR for the current
sample of HDFS galaxies we use the relationship given by Kennicutt (1998)
assuming a Salpeter IMF (0.1--100 M$_{\odot}$), a constant star formation scenario and solar abundances:

\begin{equation}
 SFR(M_{\odot}{\rm /yr}) = 1.4 \times 10^{-41} L({\rm [O II]})
({\rm erg~s}^{-1})
\end{equation}

[OII]$\lambda$3727 luminosities were calculated from the measured 
[OII]$\lambda$3727 fluxes and the measured redshifts. We applied aperture 
correction factors (with
average value of 1.3) to
account for the light missed by our 1.2'' slit. We calculated aperture
corrections in two different ways: (1) we smoothed the WFPC2 F814 image to a
resolution of 1.2 '' (to simulate the seeing during the present
observations) and then measured the ratio between the total object
counts ( in the image) and the object counts through an 
artificial 1.2'' slit; 
(2) for each galaxy we estimated
the continuum flux at 3727\AA from the $V_{AB}$
magnitude (recall that for z$\sim$0.6 the V band filter samples the rest-frame
UV light) and calculated the [OII]$\lambda$3727 flux from the 
Equivalent Width (EW, [OII]$\lambda$3727) measured from our spectra. 
Both methods lead to similar correction factors of $\sim$1.3.

However, because the observed [OII]$\lambda$3727 line flux is extremely 
sensitive to
extinction, an estimate of extinction should be made before we calculate SFRs.
We base our extinction estimates on the comparison between predicted and 
observed $V-K$ colour index (magnitudes taken from Table 2).
We use the Starburst99 code (Leitherer et al.~1999) for various star formation
histories (i.e., bursts of different durations and ages,
and continuous star formation) and calculate the range of intrinsic 
colours. The intrinsic V--K colours predicted by the models
are in the range 1.1 -- 1.5.
We apply infrared (Poggianti 1997)
and optical (Coleman 1980) K-corrections for spiral (Sc) galaxies
(assuming the median redshift of our sample z$\sim$0.6).
Comparing the observed $V-K$ colours of our galaxies {\bf with} the predicted 
ones
we obtain a median extinction $A_{\small V}$ of 0.9
assuming a screen model for the extinction (we assume no extinction in 
the K-band).
This
corresponds to an upwards correction factor of 3.5 in 
the [OII]$\lambda$3727 line
fluxes. We list the final aperture-corrected and extinction-corrected 
individual SFRs in Table 3.

%Finally, we have estimated the global SFR of the Universe as a
%function of redshift. To calculate the global SFR we summed over the
%SFRs (binned according to redshift in intervals of 0.2) listed in Table 2.
%We have estimated a value of 0.05 M$_{\odot}$ yr$^{-1}$ Mpc$^{-3}$
%at a median redshift of $<$0.6$>$ (using H$_{0}$ = 70 km s$^{-1}$
%Mpc$^{-3}$, q$_{0}$ = 0.5).

\section{[OII]$\lambda$3727 Luminosity function and {\it global} SFR density}

\subsection{[OII]$\lambda$3727 Luminosity Function}

A direct estimate of the present-day SFR activity in the Universe can
be obtained by constructing the [OII]$\lambda$3727 luminosity 
function (LF) from the present sample galaxies.
We used the 
the V$/$V$_{max}$ method (e.g. Felten 1977) to estimate the number density
of galaxies in the various luminosity bins. 
In computing V$_{max}$ however, we have to take into account that the 
sample is incomplete 
especially towards the fainter magnitude bins. Given the completeness function
(Eqn 1) estimated in Section 3.2 the appropriate formula for each galaxy's
volume V$_{max}$ is:
\begin{equation}
V_{max} = \int \eta_{i} \frac{d^{2}V_{z}}{d\Omega dz} \Delta \Omega dz
\end{equation}

In the V$/$V$_{max}$ method the luminosity density is obtained from:
\begin{equation}
\phi(L)\Delta(L) = \sum_{i}  \frac{1}{V_{max} (L_i)}
\end{equation}
where V$_{max}$ is the maximum volume over which a galaxy 
is observable given 
the survey's apparent luminosity and the selection criteria. The sum
includes all galaxies in the luminosity interval L$_{i} \pm$ 0.5$\Delta$L.
The errors are 
obtained assuming a Gaussian distribution which may not be very realistic for 
bins at the bright and faint ends where statistics are rather poor.

The LF constructed according to the V$_{max}$ method is
shown in Fig. 5. 
We then fit the LF with a Schechter (1976) function, and the
best-fitting parameters (for a limiting [OII]$\lambda$3727 flux of 
1.02 $\times$10$^{-17}$ erg cm$^{-2}$ s$^{-1}$ ) are:
$\alpha$ = -1.24$\pm$0.12, $\phi^{*}$ = 10 $^{-2.56 \pm 0.08}$ Mpc $^{-3}$,
L$^{*}$ = 10$^{42.33 \pm 0.07}$ erg s$^{-1}$.
The solid line in Fig. 5 represents the Schechter
function for the LF.

\begin{figure*}
\psfig{file=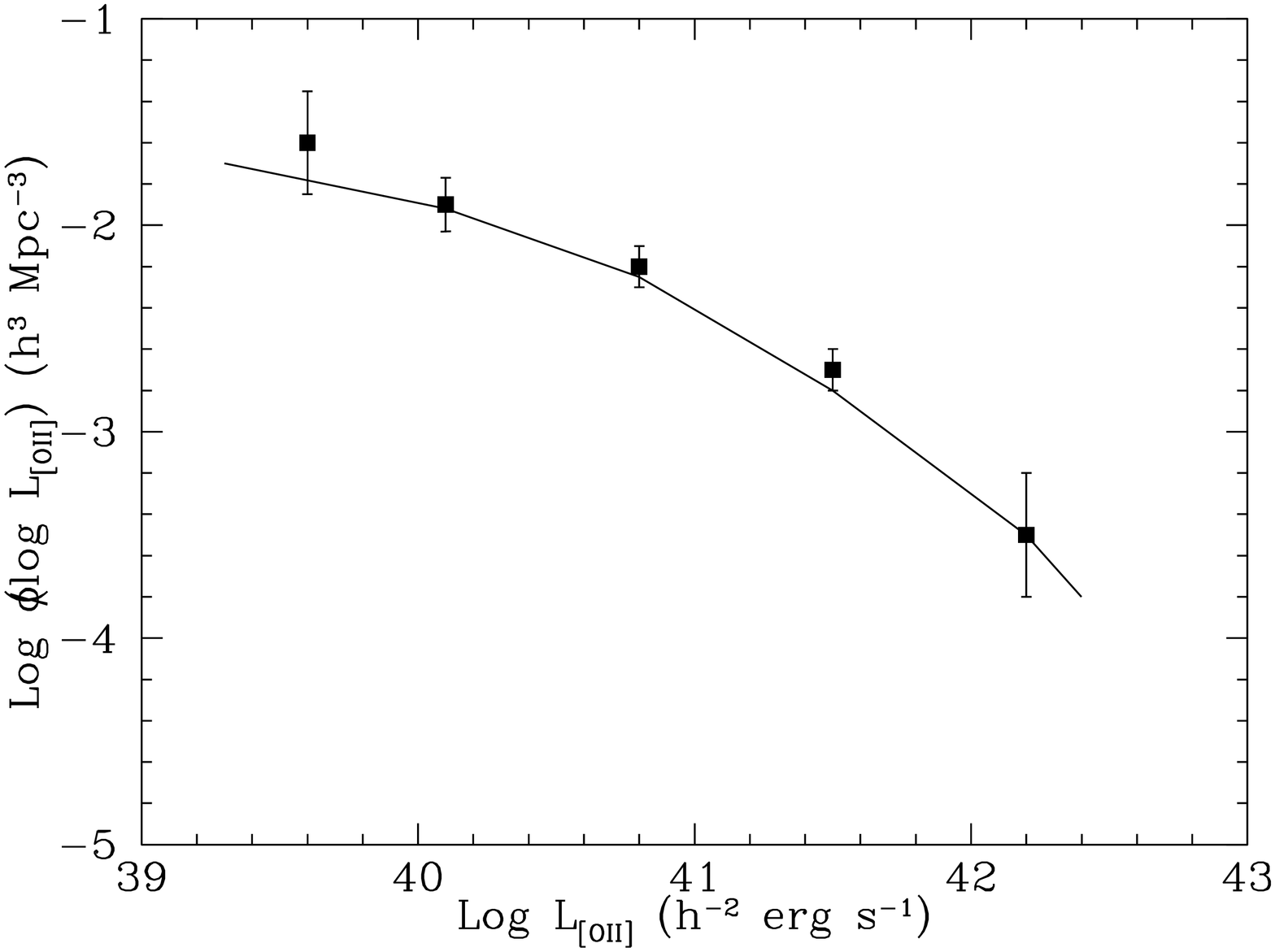,height=16.0 cm,width=16 cm,angle=-90}
\caption{[OII] luminosity function based on the V$_{max}$ method 
with no extinction-corrections applied.}
\end{figure*}

Since the luminosity function is well fitted by a Schechter
function with $\alpha \leq$-2, $\phi$(L) can be integrated over the whole
range of luminosities :
\begin{equation}
L_{tot} = \int L \phi(L)dL = \phi^{*} L^{*} \Gamma(2 + \alpha)
\end{equation}
For the observed luminosities the {\it total} [OII]$\lambda$3727 
luminosity density is 
L$_{[OII]}$ = 10$^{39.42 \pm 0.05}$ erg s$^{-1}$ Mpc$^{-3}$ 
(for 0.5$<$z$<$1.5, $<$z$>$=0.8). We note that for a Schechter function the 
luminosity density is dominated by L$_{*}$ galaxies, so galaxies outside the 
observed luminosity range do not introduce large errors as long as galaxies
near L$_{*}$ are included and the faint end slope is $\alpha <$ = -2 (e.g.
Tresse et al. 2001). A similar result for the {\it total} luminosity density 
is obtained by summing up L[OII]$/$V$_{max}$ for each galaxy. This gives the 
directly observed luminosity density 
L$_{[OII]}$ = 10$^{39.12}$ erg s$^{-1}$ Mpc$^{-3}$.

In addition to the total [OII]$\lambda$3727 luminosity density estimate
we divided our sample into three redshift bins at $<$z$>$=0.4 and $<$z$>$=0.8
and $<$z$>$=1.2, while for each individual bin we estimated the 
[OII]$\lambda$3727 luminosity density. For each individual bin we 
used the sum of the individual densities for each galaxy 
(as described above) to find:
10$^{38.40 \pm 0.19}$ erg s$^{-1}$ Mpc$^{-3}$ at $<$z$>$=0.4, 
10$^{38.70 \pm 0.14}$ erg s$^{-1}$ Mpc$^{-3}$ at $<$z$>$=0.8, and
10$^{38.90 \pm 0.22}$ erg s$^{-1}$ Mpc$^{-3}$ at $<$z$>$=1.2. 
The [OII] luminosity density as a function of median redshift is shown in 
Figure 6.
Although these
later estimates suffer from small number statistics (in each bin we have 
10$<$n$<$20 galaxies), we note an evident trend towards higher
luminosity density with increasing redshift over the range 0.4$<$z$<$1.2.

\begin{figure*}
\psfig{file=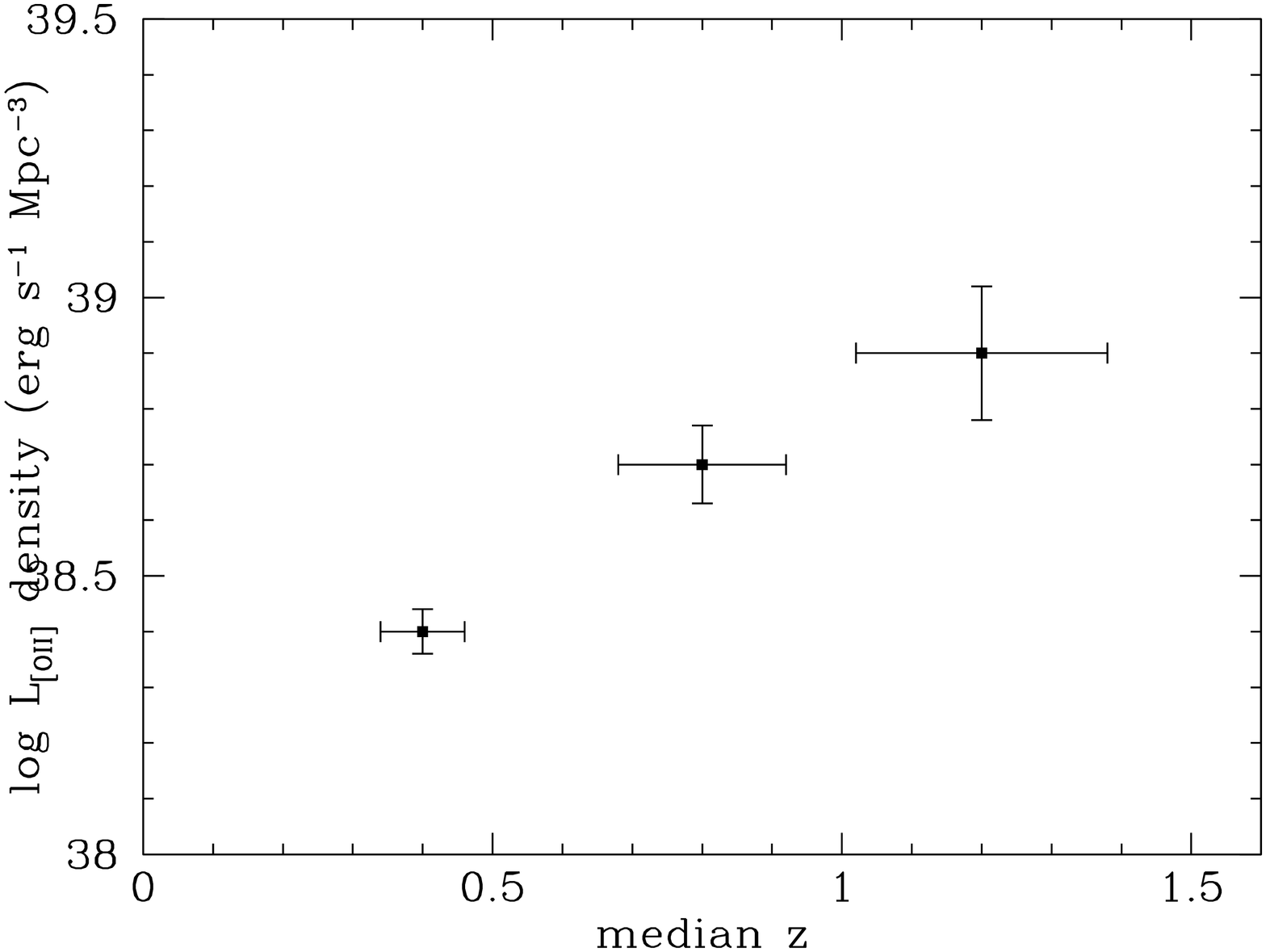,height=16.0 cm,width=16 cm,angle=-90}
\caption{[OII] luminosity density as a function of median redshift.}
\end{figure*}

\subsection{Global SFR density}

Using Equation (2)
we can transform L[OII]$\lambda$3727 into a {\it global} SFR density. 
The total [OII]$\lambda$3727 luminosity density
we estimated in 5.1 translates into an SFR density of 0.015$\pm$0.03
M$_{\odot}$ yr$^{-1}$ Mpc$^{-3}$. If we assume a median extinction of 
A$_{V}$= 0.9 (see Section 4) then the total 
extinction-corrected [OII]$\lambda$3727 
luminosity density translates into SFR density of 0.11$\pm$0.02 
M$_{\odot}$ yr$^{-1}$ Mpc$^{-3}$.

In Figure 7 we plot SFR densities vs. redshifts 
from the present spectroscopic survey (for the redshift range
0.4--1.2) supplemented with data from other surveys in the
literature. We restricted ourselves to SFR estimates based
on [OII]$\lambda$3727 surveys in order to avoid 
conversions from H$_{\alpha}$ to 
[OII]$\lambda$3727 luminosities. 
For the local Universe we used the SFR estimated by 
Gallego et al. (2002), and Hammer et al., (1997) for redshifts between 
0.5 and 1.0. For redshifts beyond z$\simeq$1.5 we used the results
from Vanzella et al. (2002) and Steidel et al. (1997). Our results are in
good agreement with the Canada-France-Redshift-Survey 
(CFRS, Hammer et al., 1997) results for the same redshift range. 
Unfortunately, most surveys provide only the observed 
luminosities and not the extinction corrected values. 
For the present data we plot both observed and 
extinction-corrected values. 

\begin{figure*}
\psfig{file=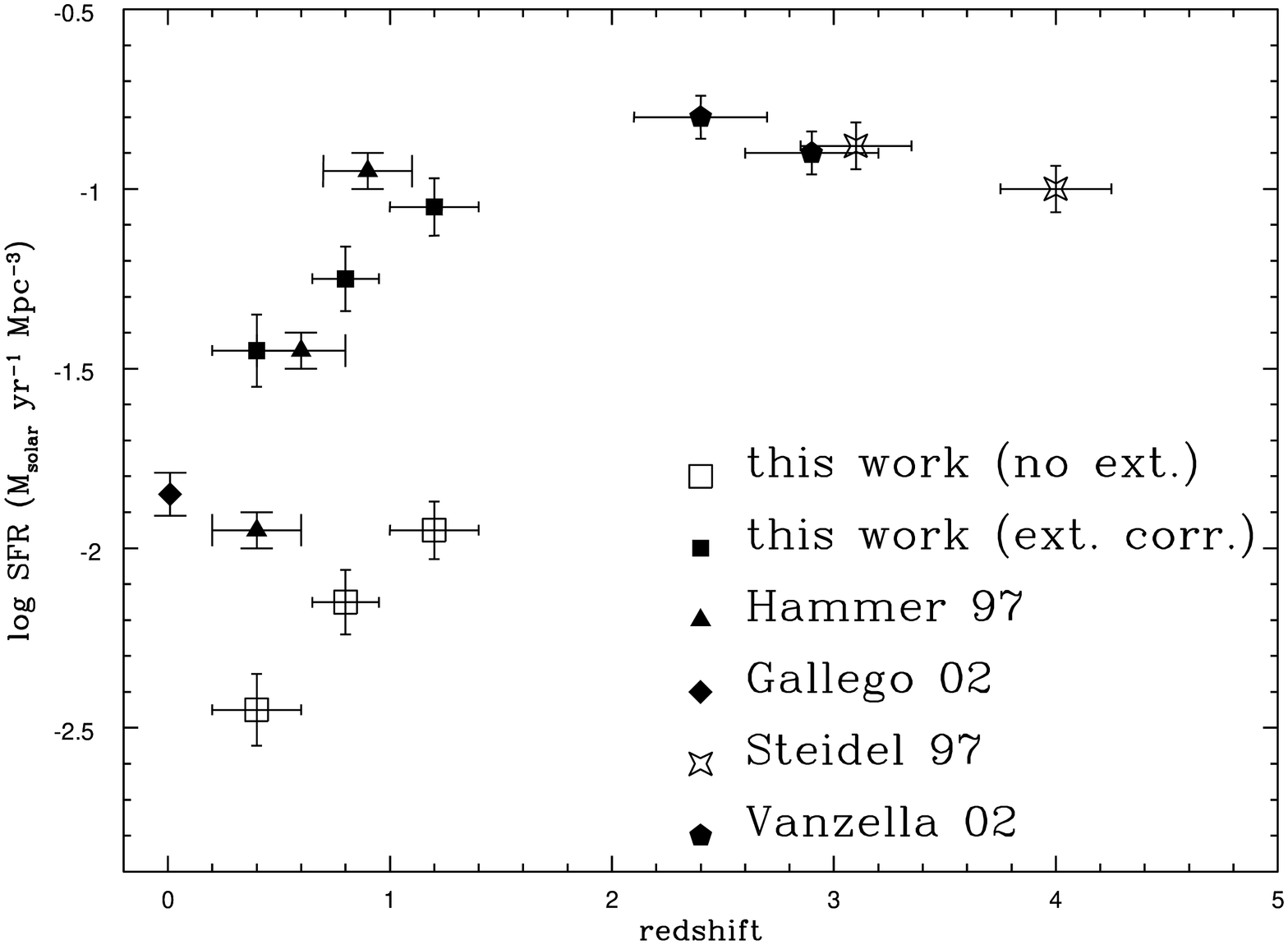,height=16.0 cm,width=16 cm,angle=-90}
\caption{SFR densities based on the present data, supplemented with literature
values. Both the observed and extinction-corrected values are shown for the 
present sample and are represented as open and filled squares. Filled diamonds correspond to Gallego et al. (2002); filled triangles are for Hammer et al. (1997), filled pentagons for Vanzella et al. (2002) and filled stars are for Steidel
et al. (1997).}
\end{figure*}
Our present data show evidence for strong evolution from the 
local Universe out to z$\simeq$1.2. Such a strong evolution of the SFR has
also been observed with the H$_{\alpha}$ survey of Tresse et al. (2002)
and the CFRS (Hammer et al. 1997). 
Assuming that the evolution of the SFR density with redshift follows
a relationship of the (1$+$z)$^n$ form, then the combination of all the SFR
data points implies a value for n$\simeq$4. We caution, though,
that this result is to some degree dependent on the local SFR density 
value used. As an example the local Universe SFR density value 
derived from the  H$_{\alpha}$ survey of Singleton et al. (2001) is lower 
than the value of Gallego et al. (1995). Irrespective of the local value, 
however, our present data show a clear rise in the SFR density by a factor
of 6 in the 0.4$<$z$<$1.2 regime.

\section{Conclusions}

We have presented new spectroscopic redshifts for 50 galaxies in the HDFS
and Flanking Fields. We determined redshifts based on the detections of
several emission lines such as [OII]$\lambda$3727, 
H$_{\beta}$ and [OIII]$\lambda$5007 and the absorption features CaII H, K 
$\lambda \lambda$ 3968.5 and 3933.7 \.A.

The redshift range of the present galaxy sample is 
0.6$<$z$<$1.2 with a median redshift of 1.13 (at I$\simeq$23.5 not 
corrected for completeness). A large fraction of the galaxies detected are 
starbursts while the remaining $\sim$10\% are ellipticals.
The {\it individual} star formation rates have been estimated for the emission
line objects and the extinction correction rates range between 
0.5--30 M$_{\odot} /$ yr.
Based on the present data we have estimated the {\it global} SFR density 
of the Universe out to z$\sim$1.3 and found it to be 0.11$\pm$0.02 
M$_{\odot}$ yr$^{-1}$ Mpc$^{-3}$ using the total extinction 
corrected [OII]$\lambda$3727 luminosity density.

Our data show evidence of a strong
evolution from the local (z=0) Universe out to z$\sim$1.3, following 
(1$+$z)$^{4}$. Finally, the
availability of 8m class telescopes allows us to sample
the evolution of the z$\simeq$1 Universe precisely and improve on previous 
determinations. 

\begin{acknowledgements}

We acknowledge support of the EU TMR Network ``Probing the Origin of
the Extragalactic  Background Radiation'' (HPRN-CT-2000-00138). We thank
Michel Dennefeld and his team for acquiring the data and Niranjan Thatte
for help with data analysis. SB thanks MPE for the hospitality during his
visits.
\end{acknowledgements}


\begin{thebibliography}{}

\bibitem[Appenzeller 2000]{app2000}
Appenzeller, I., Bender, R., Boehnhardt, H., et al.\ 2000, ESO
Messenger, 100, 44

\bibitem[Arnouts 2002]{arn2002}
Arnouts, S., Moscardini, L., Vanzella, E., et al.\ 2002, MNRAS 329, 355

\bibitem[Bergeron 1999]{berg1999}
Bergeron, J., Petitjean, P., Cristiani, S., et al.\ 1999, \aap 343, 40

\bibitem[Casertano 2000]{cas2000}
Casertano, S., et al.\ 2000, \aj, 120, 2747

\bibitem[Cohen 2000]{cohen2000}
Cohen, J.G., Hogg, D.W., Blandford, R., et al. \ 2000, \apj 538, 29 

\bibitem[Coleman 1980]{coleman80}
Coleman, G.D., Wu, C.-C., Weedman, D.W. \ 1980, \apj 43, 393

\bibitem[Cristiani 2000]{cris2000}
Cristiani, S., Appenzeller, I., Arnouts, S., et al. \ 2000, \aap, 359, 489

\bibitem[da Costa 1998]{dac98}
Da Costa, L., et al.\ 1998, \aap submitted, astroph$/$9812105

\bibitem[Felten 1977]{felten77}
Felten, J.E., \ 1977, \aj 82, 861

\bibitem[Franceschini 2003]{fra2003}
Franceschini, A., Berta, S., Rigopoulou, D., et al. \ 2003, \aap, 403, 501

\bibitem[Gallego 1995]{gal1995}
Gallego, J., Zamorano, J., Aragon-Salamanca, A., 
Rego, M. \ 1995, \apj Lett 455, 1

\bibitem[Gallego 2002]{gal2002}
Gallego, J., Garcia-Dabo, C.E.,Zamorano, J., Aragon-Salamanca, A., Rego, M.
\ 2002, \apj Lett, 570, 1

\bibitem[Hammer 1997]{ham97}
Hammer, F., Flores, H., Lilly, S.J., \ 1997, \apj 481, 49

\bibitem[Kennicutt 1998]{kenn98}
Kennicutt, R.C., \ 1998, ARAA, 36, 189

\bibitem[Leitherer 1999]{leith99}
Leitherer, C., et al., \ 1999, ApJS, 123, 3

\bibitem[Mann 2002]{mann2002}
Mann, R.G., Oliver, S., Carballo, R., et al. \ 2002, MNRAS, 332, 549

\bibitem[Oliver 2002]{oli2002}
Oliver, S., Mann, R.G., Carballo, R., et al.\ 2002, MNRAS, 332, 536

\bibitem[Page 2000]{page2000}
Page, M.J., \& Carrera, F.J., \ 2000, MNRAS, 311, 433

\bibitem[Poggianti 1997]{pog97}
Poggianti, B.\ M.\ 1997, AAS, 122, 399

\bibitem[Rigopoulou 2000]{rig2000}
Rigopoulou, D., et al.\ 2000, ApJ, 537, L85

\bibitem[Rudnick 2001]{rud2001}
Rudnick, G., et al.\ 2001, AJ, 122, 2205

\bibitem[Saracco 2001]{sar2001}
Saracco, P., Giallongo, E., Cristiani, S., D'Odorico, S., Fontana, A., Iovino, A.,Poli, F., \& Vanzella, E.\ 2001, \aap, 375, 1

\bibitem[Sawicki 2003]{saw2003}
Sawicki, M., \& Mallen-Ornellas, G., 2003, \aj, 126, 1208


\bibitem[Schechter 1976]{sche1976}
Schechter, P., \ 1976, \apj 203, 297

\bibitem[Singleton 2001]{sing2001}
Singleton, C., \ 2001, PhD thesis, Univ. Nottingham


%\bibitem[Sternberg 1998]{stern98}
%Sternberg, A.,\ 1998, \apj, 506, 721

\bibitem[Teplitz 2001]{tep2001}
Teplitz, H.\ I., Hill, R.\ S., Malumuth, E.\ M., Collins, N.\ R., Gardner, J.\ P.,
Palunas, P., \& Woodgate, B.\ E.\ 2001, ApJ, 548, 127

\bibitem[Tresse 1999]{tres99}
Tresse, L., Dennefeld, M., Petitjean, P., Cristiani, S., White, S., et
al.\ 1999, \aap 346, 21

\bibitem[Vanzella 2001]{vanz2001}
Vanzella, E., Cristiani, S., Saracco, P., Arnouts, S., 
et al.\ 2001, AJ, 122, 2190

\bibitem[Vanzella 2002]{vanz2002}
Vanzella, E., Cristiani, S., Arnouts, S., et al.\ 2002, \aap, 396, 847

\bibitem[Williams 2000]{will2000}
Williams, R.\ E., et al.\ 2000, \aj, 120, 2735

\end{thebibliography}
\end{document}